\documentstyle[preprint,aps]{revtex}
\begin{document}
%\draft

\title{Determination of the $\pi NN$ Form Factor from the Threshold Pion
Production of
$pp$ Scattering}
\author{T.-S.\ H.\ Lee}
\address{Physics Division, Argonne National Laboratory, Argonne, IL
60439-4843 USA}

\maketitle
\begin{abstract}
It is shown that the threshold productions of $\pi^0 pp$, $\pi^+ np$ and $\pi^+
d$
from $pp$ collisions can be consistently described by a model consisting of a
pion s-wave
rescattering and $N\bar N$ pair-terms of
heavy-meson exchanges.  The large difference between $\sigma^{tot}
(pp\rightarrow\pi ^+ d)$
and $\sigma^{tot} (pp\rightarrow\pi ^+ np)$
is understood from the orthogonality of the deuteron and the $np$ scattering
wave functions.
In a calculation using the Paris potential to account for the initial and
final $NN$ interactions, it is found that the data can be best reproduced
by using a soft $\pi NN$ form factor with $\Lambda_{\pi} =650 $ MeV
for a monopole form.
This is consistent with an earlier study of pion production in the $\Delta$
excitation region.

\end{abstract}

%\narrowtext

\newpage

With the development of the Cooler facility at the Indiana University
Cyclotron Facility (IUCF), accurate data of threshold pion production from
proton-proton ($pp$) scattering have become available.  The data of
$pp \rightarrow pp \pi^0$ \cite{meyer90} had led to the
identification of the short-range
pion production mechanisms \cite{lee93,niskanen94,horowitz94}.
In this paper we demonstrate that by also including
the new data of the $pp \rightarrow \pi^+ np$ \cite{daehnick} and $pp
\rightarrow \pi^+ d$
\cite{heimberg94}
reactions in a theoretical analysis, we can extract information concerning
the range of the $\pi NN$ form factor.  Furthermore, we will show that the
large
difference between the cross sections of $pp \rightarrow \pi^+ d$ and
$pp \rightarrow \pi^+ np$ can be understood from the behavior of
the final $np$ wave functions.

Following the previous
investigations\cite{lee93,niskanen94,horowitz94,koltun66,miller91}, the
s-matrix for the
production of a pion with a charge $\alpha$ from a $pp$ collision
can be written as (we use the normalization of \cite{goldberg75})

\begin{equation}
S_{fi} = -2\pi i \delta^{(4)} (p_1 + p_2 - p_f - k) \langle f|A_{\alpha} |
\chi ^{(+)}_{\vec p_{1} \vec p_{2}} \rangle
\end{equation}
where $|\chi^{(+)}_{\vec p_{1} \vec p_{2}} \rangle$ is the initial $pp$
scattering wave
funtion, and $\langle f|$ is either the deuteron wave function
in $pp \rightarrow \pi^{+} d$ reaction or the
$np$ scattering wave function
$\langle \chi ^{(-)}_{\vec p'_{1} \vec {p}'_{2}}|$ in
$pp \rightarrow \pi^{+} np$ reaction.  Following the previous approach
\cite{lee93},
the plane-wave matrix element of the
production operator $A_{\alpha}$ is defined by taking a nonrelativistic limit
of
the lowest-order Feynman amplitudes
calculated from a model interaction Hamiltonian.  For our present
purpose, it is sufficient to consider the following interaction Hamiltonians
for
$\pi$, $\omega$, $\sigma$, and $N$ fields

\begin{equation}
H(x) = H_{\pi} (x) + H_s + H_{\sigma} (x) + H_{\omega} (x)
\end{equation}
where

\begin{eqnarray}
H_{\pi} (x) &=& {f \over m_{\pi}} \,\bar{\psi} (x) \gamma_5
\gamma^{\mu}\,\vec{\tau}\,
\psi (x) \cdot \partial_\mu \vec{\phi}_{\pi} (x)
\nonumber \\
H_s (x) &=& {4\pi \lambda_1 \over m_{\pi}}\,\bar{\psi} (x) \vec{\phi}_{\pi} (x)
\cdot \vec{\phi}_{\pi} (x) \psi (x)
+ {4\pi \lambda_2 \over m_{\pi}^{2}}\,\bar{\psi} (x) \vec{\tau}\,\psi  (x)
\cdot
\vec{\phi} (x) \times \left({\partial \over \partial t} \vec{\phi}\,(x) \right)
\nonumber \\
H_{\sigma} (x) &=& g_{\sigma} \bar{\psi} (x) \psi (x) \phi_{\sigma} (x)
\nonumber \\
H_{\omega} (x) &=& g_{\omega} \bar{\psi} (x) \gamma^{\mu} \psi (x)
\partial_{\mu}
\phi_{\omega} (x)
\end{eqnarray}
Here $H_{\pi}$ describes the familiar pseudovector $\pi NN$ coupling.  The $\pi
N$ s-wave
interaction is described by $H_s$, as introduced in Ref.\  \cite{koltun66}.
In this work, we use the vaules $\lambda_1 \simeq 0.005$
and $\lambda_2 \simeq 0.05$
calculated\cite{lee93} from the $\pi N$ s-wave scattering lengths
of \cite{hohler79}.
The heavy meson-exchanges are induced by $H_{\sigma}$ for a isoscalar-scalar
$\sigma$
meson and $H_{\omega}$ for a isoscalar-vector $\omega$ meson.  The isovector
mesons
$\delta$ and $\rho$ can be included \cite{lee93}, but will be neglected in this
work for
simplicity.

The pion s-wave rescattering term of the production operator $A_{\alpha}$
is calculated from $H_{\pi}$ and $H_s$.
The $N\bar N$ pair-terms of $\sigma$ and $\omega$ exchange amplitudes are
calculated from
$H_{\pi}$ and second orders in
$H_{\sigma}$ and $H_{\omega}$. To ensure that there is no double counting in
the
calculation of Eq.\ (1), only the negative energy component of the nucleon
propagaor
in the heavy-meson Feynman amplitudes is kept.  The resulting terms are called
the
$N\bar N$ pair-terms.
To be consistent with the employed nonrelativistic $NN$ potentials for
calculating
the initial and final $NN$ wave functions, we make the usual nonrelativistic
expansion of the Feynman amplitudes.
In terms of momentum variables and pion isospin
($N(p_1) + N(p_2) \rightarrow N(p'_1) + N(p'_2 )+ \pi (k,\alpha ))$,
we obtain the following form

\begin{equation}
A_{\alpha} = \left[ T^{(1)}_{\alpha} +
T^{(\pi )}_{\alpha} + T^{(\sigma )}_{\alpha} + T^{(\omega)}_{\alpha} \right]
+ ( 1 \longleftrightarrow 2)
\end{equation}
where

\begin{eqnarray}
T^{(1)}_{\alpha} &=& G_0 \left\{ \delta (\vec{p}_2 - \vec{p'}_2 ) (2\pi)^3
\left[ \vec{\sigma}\,(1)
\cdot \vec{k} - {\omega _k \over 2m_N} \,\vec{\sigma}\,(1) \cdot
(\vec{p}_1 + \vec{p'}_1) \right] (-1)^{\alpha} \tau_{-\alpha} (1) \right\}
\nonumber \\
T^{(\pi)}_{\alpha} &=& G^{(\pi)} \left\{ {\vec{\sigma}_2 \cdot (\vec{p}_2 -
\vec{p'}_2) \over
-\vec q\,^{2} - m^{2}_{\pi}} \left[ {8\pi \lambda_{1} \over m_{\pi}}
(-1)^{\alpha} \tau_{-\alpha} (2)
+ i\, {8\pi \lambda_{2} \over m_{\pi}}\, {\omega _k + \omega _q \over 2m_{\pi}}
 (-1)^{\alpha} \left[ \vec{\tau} (1) \times \vec{\tau} (2) \right] _{- \alpha}
\right] \right\}
\nonumber \\
T^{(\sigma )}_{\alpha} &=& G^{(\sigma )} \left\{ {\omega _{k} \over -\vec
q\,^{2} - m^{2}
_{\sigma}}\,{g^{2}_{\sigma} \over m^{2}_{N}}\,
{\vec{\sigma} (1) \cdot (\vec{p}_1 + \vec{p'}_1 + \vec{k})} (-1)^{\alpha}
\tau_{-\alpha}
(1) \right\}
\nonumber \\
T^{(\omega)}_{\alpha} &=& G^{(\omega)} \left\{ {\omega _k \over -\vec q\,^{2} -
m^{2}_{\omega}}
{-g^{2}_{\omega} \over m^{2}_{N}}
\left[ {\vec{\sigma} (1) \cdot (\vec{p}_2 + \vec{p'}_2)} + {i \over 2}
(\vec{p'}_2 - \vec{p}_2 )
\cdot (\vec{\sigma} (1) \times \vec{\sigma} (2)) \right] (-1)^{\alpha}
\tau_{-\alpha} (1) \right\}
\end{eqnarray}
with

\begin{eqnarray}
&&\left\{
\begin{array}{cc}
G_0 = {i \over (2\pi)^{9/2}}\, {1 \over \sqrt{2\omega _k}}\, {f \over
m_{\pi}}\,,\\
G^{(i)} = G_0 \left[ F^{(i)} (\Lambda _i , \vec q\,^2 ) \right]^2\,.
\end{array}\right.
\end{eqnarray}
Here we set $\vec q = \vec p _2 - \vec p\,'_{2}$\,, $\omega _k =
\sqrt{m^{2}_{\pi} + k^2}$\,,
$\omega _q = \sqrt{m^{2}_{\pi} + q^2}$,
$\tau_{\pm} = {\mp \over \sqrt{2}}\,
(\tau_1 \pm i\, \tau_2 )$, $\tau_0 = \tau_z$,
and $\alpha$ is the z-component of pion isospin.
The form factors $F^{(i)}(\Lambda_{i} ,
\vec q\,^{2} )$ are introduced to regularize meson-NN vertices.

Let us first consider the pion s-wave rescattering term $T^{\pi}$.  Its main
feature is that
$\lambda _1 \simeq 0.005 \ll \lambda _2 \simeq 0.05$.  Since the matrix element
of the
isospin operator
($\vec \tau _1 \times \vec \tau _2$) between two $T = 1$ states vanishes, only
the very weak
$\lambda _1$ term of $T^{\pi}$ can have a contribution to the $pp \rightarrow
pp \pi ^0$
cross section.  That is why the calculations
\cite{lee93,niskanen94,horowitz94,miller91} including only $T^{(1)} +
T^{(\pi)}$
failed to account for the total cross section of $pp \rightarrow pp \pi ^0$.
It was found \cite{lee93,niskanen94,horowitz94} that the discrepancies with the
data can
be removed by
introducing the $N\bar N$ pair terms of heavy-meson exchange.  A more general
approach
was introduced in Ref.\ \cite{lee93} to calculate the $N\bar N$ pair-terms from
realistic
$NN$ potentials.  For the present purpose,
it is sufficient to use the simple model defined in Eqs.\ (5). We choose the
monopole form
factor
$F^{(i)} (\Lambda _{i} , \vec q\,^{2} ) = \left(
{\Lambda ^{2}_{i} - m^{2}_{i} \over \Lambda ^{2}_{i} + \vec q\,^{2}} \right)$.
All of the two nucleon wave functions in Eq.\ (1) are calculated from the Paris
potential \cite{lacombe80}.  We found that the threshold
$pp \rightarrow pp \pi ^0$ total cross section data can be best reproduced if
we choose
$g_{\sigma}^{2}/4\pi = 6$ and $\Lambda_{\sigma} = 1000$ MeV/c, for the
$\sigma$-exchange,
and $g_{\omega}^{2}/4\pi = 10$ and $\Lambda_{\omega} = 1400$ MeV for the
$\omega$-exchange.
In considering only the total cross sections, one can also fit the data by
keeping only the
$\sigma$-exchange term with $g_{\sigma}^{2}/4\pi = 12$ and $\Lambda_{\sigma} =
2000$ MeV/c.
This is similar to the findings of Refs. \cite{niskanen94} and
\cite{horowitz94}.
All results presented below are from the $\sigma$-exchange model.  The results
for
$\sigma + \omega$-exchange model are very similar and are therefore omitted.

Once the heavy meson-exchange mechanism has been determined in the study of the
$pp \rightarrow pp \pi ^0$ reaction, the only freedom of the model is the
form factor $F^{(\pi )} (\Lambda_{\pi} , \vec{q}\,^{2} )$
of $T^{(\pi)}$.  Since the $T = 0$ final $np$ states are involved in
the charged pion production, the large $\lambda_2$ term of $T^{(\pi)}$ will
contribute
to the cross section.  Its contribution depends on the range $\Lambda _{\pi}$
of the
$\pi NN$ form factor.
We find that all of the threshold pion data can be best reproduced with
$\Lambda_{\pi} = 650$ MeV/c.  Our results are the solid curves displayed in
Fig.\ 1.  In
the same figure, we also show the results (dotted curves) without including the
$N\bar N$
pair-terms of heavy
meson-exchanges.  The heavy meson-exchange mechanisms clearly dominate the
$pp \rightarrow pp \pi ^0$
cross section; but also plays a significant role in charged pion productions.

An interesting feature of Fig.\ 1 is that at energies near the $\pi ^+ np$
production
threshold (292.30 MeV), the cross sections of the $\pi ^+ d$ production are
almost two
orders of magnitude larger than the $\pi ^+ np$ production.  This can be
understood as
follows.  In this near threshold energy region, the final $np$ pair
is mainly in the
$^1 S _0 (T = 1)$ and $^3 S -\,^3 D_1 (T = 0)$ states.  The production of the
$^1 S_0 (T = 1) np$ state is very weak mainly because the large $\lambda _2$
term of the
pion rescattering $T^{(\pi)}$ does not contribute (as in the case of $pp
\rightarrow pp \pi ^0$).
The difference between the $\pi ^+ d$ and $\pi ^+ np$ productions is therefore
only in their
final radial wave functions of the $np$ system.  With the same $^3 S_1 -\,^3
D_1$ quantum
number, the $np$ scattering wave function should be orthogonal to the deuteron
wave
function.  As illustrated in Fig.\ 2, this comes about from any realistic $NN$
potential.
The orthogonality is due to the change of the sign of the $np$ scattering wave
function
in the region where the deuteron wave function is large.  Consequently, the
$\pi ^+ np$
production amplitude involves a cancellation in the integration over the
production
operator, and hence is very much suppressed.  The results shown in Fig.~1
reflect the
correctness of the employed Paris potential in describing this wave function
effect.

We have found that our predictions are very sensitive to the range $\Lambda
_{\pi}$
of the form factor $F^{(\pi)}$.  This is not surprising since the threshold
pion production
involves a large momentum
transfer at the $\pi NN$ vertex.  In Fig.\ 3, we show that the calculations
with $\Lambda_{\pi} = 1000$ MeV/c overestimate significantly the charged pion
production
cross sections.  This is consistent with the results from the earlier work
\cite{lee83}
in the $\Delta$-excitation region.

The results presented above are from the calculations using the $\pi NN$
coupling
constant $f^2/4\pi = 0.079$ of Ref.\cite{hohler79}. In recent $\pi
N$\cite{arndt85} and $NN$\cite{stoks93,arndt94} phase shift analyses, a smaller
value of $f^2/4\pi = 0.075$ was obtained. We have also carried out calculations
using this smaller $\pi NN$ coupling constant. A slightly larger $\pi NN$
cutoff $\Lambda = 680$ MeV is needed to obtain an equally good fit
to all of the data shown in Figs.1 and 3.

In conclusion, we have shown that all of the threshold pion production data
from $pp$
collisions can be described by a model consisting of the pion s-wave
rescattering term
and the $N\bar N$ pair terms of heavy meson-exchanges.  The large differences
between the
$\pi ^+ d$ and $\pi ^+ np$ production are shown to be due to the orthogonality
between the
$np$ scattering wave function and the deuteron bound state.  Because of the
isospin
character of the pion rescattering term, the $pp \rightarrow pp \pi ^0$ data
can be used
to pin down the $N\bar N$ pair terms of the heavy meson-exchanges.  The charged
pion
production can therefore be used to determine the $\pi NN$ form factor.  A soft
$\pi NN$ form factor with $\Lambda_{\pi} = 650$ MeV for a monopole form is
found to be
consistent with the data. To end, we mention that the present investigation
is based on the distorted-wave formulation, Eq.(1), which was used in all of
the previous studies\cite{lee93,niskanen94,horowitz94,koltun66,miller91}.
To make further progress, it is necessary to consider the coupling between
the final $\pi^0 pp$, $\pi^+ np$ and $\pi^+d$ channels. This is a highly
nontrivial task. Our effort in this direction will be reported elsewhere.

I would like to thank Steve Dytman and Ralph Segel for informing me of their
data, Vincent Stoks and Iraj Afnan for useful discussions.
This work is  supported by the U.S. Department of Energy, Nuclear Physics
Division, under
contract W-31-109-ENG-38.

\newpage

\begin{figure}
\caption{The total cross sections (solid curves) calculated from using the
Paris
potential and $g_{\sigma}^{2}/4\pi = 16$, $\Lambda_{\sigma} = 2000$ MeV/c for
the $\sigma$-exchange model of $A_{\alpha}$ are compared with the data.
The dotted curves do not include the $N\bar N$ pair-terms of heavy meson
exchanges.}
\end{figure}

\begin{figure}
\caption{The solid curve is the $^3 S_1$ deuteron bound state wave function.
The dotted and dashed curves are the $^3 S_1$ scattering wave function for
$E_L = 1$ and 10 MeV.  The wave functions are generated from the Paris
potential.}
\end{figure}

\begin{figure}
\caption{The solid (dotted) curves are the calculated total cross section using
the
cutoff $\Lambda _{\pi} = 650\,(1000)$ MeV for the $\pi NN$ form factor.  The
results are
from calculations using the Paris
potential and $g_{\sigma}^{2}/4\pi = 16$, $\Lambda_{\sigma} = 2000$ MeV/c for
the $\sigma$-exchange model of $A_{\alpha}$.}
\end{figure}


\newpage

\begin{references}
\bibitem{meyer90} H.\ O.\ Meyer $\it et\,al$., Phys.\ Rev.\ Lett.\ $\bf 65$,
																		2846 (1990); Nucl.\ Phys.\ $\bf A539$, 633 (1992).
\bibitem{lee93}			T.-S.\ H.\ Lee and D. O. Riska, Phys.\ Rev.\ Lett. $\bf 70$,
																		2237 (1993).
\bibitem{niskanen94}	J.\ A.\ Niskanen, Phys.\ Rev.\ C $\bf 49$, 1285 (1994).
\bibitem{horowitz94}  C.\ J.\ Horowitz, H.\ O.\ Meyer, and D.\ K.\ Griegel,
																						Phys.\ Rev. C $\bf 49$, 1337 (1994).
\bibitem{daehnick}	W.\ W.\ Daehnick, S.\ A.\ Dytman, J.\ G.\ Hardie, W.\ K.\
Brooks,
																			R.\ W.\ Flammang, L.~Bland, W.\ W.\ Jacobs, T.\ Rinckel,
																			P.\ V.\ Pancella, J.\ D.\ Brown, and E.\ Jacobson,
																			submitted to Phys.\ Rev. Lett.
\bibitem{heimberg94}		P. Heimberg $\it et\,al$., Proceedings of SPIN94
Conference,
																					 Bloomington, IN (1994) in press; D. A. Hatcheon $\it
et\,al$.,
																						Nucl.\ Phys.\ $\bf A535$, 618 (1993);
																      C. M. Kose, Jr., Phys. Rev. $\bf 154$, 1305 (1957);
		                    B. F. Ritchie $\it et\,al$., Phys.\ Rev.\ Lett. $\bf 66$,
568 (1991).

\bibitem{koltun66}	D.\ Koltun and A.\ Reitain, Phys.\ Rev.\ $\bf 141$, 1413
(1966).
\bibitem{miller91}  G.A. Miller and P.U. Sauer, Phys.\ Rev.\ C $\bf 44$,
      R1725 (1991).

\bibitem{goldberg75} M. L. Goldberger and K. L. Watson, $\it Collision\ Theory$
(Krieger, New York, 1975)
\bibitem{hohler79} G.\ H$\ddot{o}$hler $\it et\,al$., Handbook of Pion-Nucleon
																			Scattering, Physics Data $\bf 12-1$, (Karlsruhe, 1979).
\bibitem{lacombe80}     M. LaCombe $\it et\,al$., Phys.\ Rev.\ C $\bf 21$, 861
(1980).
\bibitem{lee83}  T.-S. H. Lee, Phys.\ Rev.\ Lett. $\bf 50$, 157 (1983);
             A. Matsuyama and T.-S. H. Lee, Phys. Rev. $\bf 34$, 1900 (1986)

\bibitem{arndt85} R.A. Arndt, J.M. Ford, and L.D. Roper, Phys.\ Rev.\ D
               $\bf 32$, 1085(1985).

\bibitem{stoks93} V.G.J. Stoks, R.A.M. Klomp, M.C.M. Rentmeester,
           and J.J. de Swart, Phys.\ Rev.\ C  $\bf 48$, 792(1993).

\bibitem{arndt94}R.A. Arndt, I.I. Strakosky, and R.L. Workman,
           Phys.\ Rev.\ C $\bf 50$, 2731(1994).

\end{references}
\end{document}